# Design space exploration of a poultry fillet processing system using discrete-event simulation

N. Paape[1,*], J.A.W.M. van Eekelen[1] and M.A. Reniers[1]

[1]Eindhoven University of Technology, Eindhoven, The Netherlands

*Corresponding author. Email address: n.paape@tue.nl

**Abstract**

Developments in the poultry processing industry, such as how livestock is raised and how consumers buy meat, make it increasingly difficult to design poultry processing systems that meet evolving standards. More and more iterations of (re)design are required to optimize the product flow in these systems. This paper presents a method for design space exploration of production systems using discrete-event simulation. This method automates most steps of design space exploration: iterating on the design, model construction, performing simulation experiments, and interpreting the simulation results. This greatly reduces the time and effort required to iterate through different designs. A case study is presented which shows that this method can be effective for design space exploration of poultry processing systems.

## 1. Introduction

### 1.1. Problem statement

The meat processing industry is facing rapid developments, both in how livestock is raised (Thornton, 2010), and in how consumers buy meat (Sanchez-Sabate and Sabaté, 2019). These developments are causing a mismatch between provider and consumer in the poultry industry; broilers (chickens bred for meat consumption) are being bred increasingly heavier (Verbeke and Viaene, 2000), while consumers prefer lighter chicken fillets – which they perceive as more sustainable.

This complicates the design process of the fillet processing systems in poultry processing plants in which predominantly heavy fillets need to be used for production orders which require mostly light fillets. As a result it is becoming more and more important to optimize product flows during the design process of the fillet processing system. A recent development to help combat the mismatch between inflow and orders is to introduce new machines to the fillet processing system which can trim down heavy fillets to lighter fillets. However, this also increases the number of system parameters that the system designer needs to tune (the system's design space), such as which machines are used and how these machines are connected.

Exploring the design space of poultry systems – and of production systems in general – is often an intricate process, with many iterations of (re)design. Using simulation for design space exploration of production systems allows design alternatives to be evaluated and compared (Owens and Levary, 2002). However, this generally requires the system designer to specify a design, to construct a model of this design, to perform simulation experiments on the model, and to interpret the simulation results, only to repeat these steps in the next design iteration. In this paper we aim to automate this process, not only for poultry processing systems, but for production systems in general.

### 1.2. State of the art

In this section a state of the art is given on using simulation in the design of food processing systems, and on methods which use discrete-event simulation to aid in the design space exploration of production systems in general.

### 1.2.1. Simulation for food processing systems

In Penazzi et al. (2017) simulation is used to analyze the design and control of food job-shop processing systems. A real case study from the catering industry is analyzed to identify the bottlenecks and to investigate the flow of products through the system.

In Plà-Aragonés et al. (2017, 2020), discrete-event simulation is used to compare processing alternatives and production planning for a pig meat packing plant. Plà-Aragonés et al. note that discrete-event simulation better captures the behaviour of the plant compared to deterministic or stationary approaches.

In Xie and Li (2012) an analytical model is used to identify design improvements which increase the throughput of a meat shaving and packing line.

In Owens and Levary (2002) specific design alternatives for an extruded food production line are compared using simulation of system dynamics models. The study shows that simulation can be an effective decision support tool in the food industry which can help in choosing between design alternatives.

### 1.2.2. Using simulation for design space exploration

Kikolski (2017) shows how discrete-event simulation can be used to asses the performance of a production system for a selection of production scenarios.

In Centobelli et al. (2016) a digital factory is realized in the discrete-event simulation tool Simio (Simio, 2022). This digital factory is used to optimize the facility layout with respect to the product flow. An 'as is' layout is compared to a new layout. Multiple scenarios are simulated according to different company orders.

Kranz et al. (2021) presents an algorithm which can be used to generate a (discrete-event simulation) model of an assembly system layout, based on layout data defined in an Excel spreadsheet and predefined process logic. Using this algorithm allows for the creation of simulation models of various different layouts without the need for detailed simulation expertise.

In Laemmle and Gust (2019) a method is developed for automatically generating a simulation model of a robotic cell layout based on layout data specified in AutomationML.

Rodič and Kanduč (2015) uses discrete-event simulation to optimize the facility layout of a specialized furniture manufacturing plant. In this study, the (spatial) placement of machines is optimized with respect to the total distance that products need to travel on the factory floor. The models for the different designs of the factory floor are generated automatically.

### 1.3. Contribution

The goal of this paper is to develop a method for automated design space exploration, and to use this method to optimize the functional layout of a poultry fillet processing system with respect to which machines are used and what the connections are between them. In the proposed method:

- The design space of the proposed system is explored iteratively & automatically.
- The (discrete-event simulation) models for these layouts are automatically constructed using a model library in Anylogic (The Anylogic Company, 2022).
- These models are used to carry out simulation experiments to predict the system's performance in a predefined set of production scenarios (a product scenario describes the conditions under which the production system operates).
- The system designs are then evaluated based on their performance across these production scenarios.

**Table 1.** The proposed method compared to the state of the art.

| | Kikolski (2017) | Centobelli et al. (2016) | Kranz et al. (2021) | Laemmle and Gust (2019) | Rodič and Kanduč (2015) | This paper |
|---|---|---|---|---|---|---|
| Discrete-event simulation | ✓ | ✓ | ✓ | ✓ | ✓ | ✓ |
| Comparison of design alternatives | | ✓ | ✓ | | ✓ | ✓ |
| Comparison across multiple scenarios | ✓ | ✓ | | | | ✓ |
| Automated design space exploration | | | | | ✓[1] | ✓[2] |
| Automated model construction | | | ✓ | ✓ | ✓ | ✓ |

[1] In (Rodič and Kanduč, 2015) the design space concerns the spatial layout of the production system.
[2] In this paper the design space concerns the functional layout of the production system.

The novelty of our contribution is in how these individual steps are combined to create an automated method for design space exploration, and in demonstrating how this proposed method can be applied to a case study in the poultry processing industry. Table 1 shows a comparison of the proposed method to the state of the art.

The structure of this paper is as follows: in Section 2 a case study is described which will be used to explain the proposed method for design space exploration in Section 3. Then, in Section 4 the method will be discussed, along with any revelations which were made when applying it to the case study. Finally, in Section 5 concluding remarks are made, along with ideas for future work.

## 2. Case study description

In this paper a case study is carried out to demonstrate how the proposed design space exploration method described in Section 3 can be utilized. In the case study a simplified adaptation of a real poultry fillet processing system is designed. The real system uses the same physical com-ponents, although in a different setup. It also allows for a more diverse range of order types with multiple trimming strategies. Finally, it utilizes a more elaborate production control strategy for more optimal performance. Nonetheless, the design method proposed in this paper has also been used to redesign the real system.

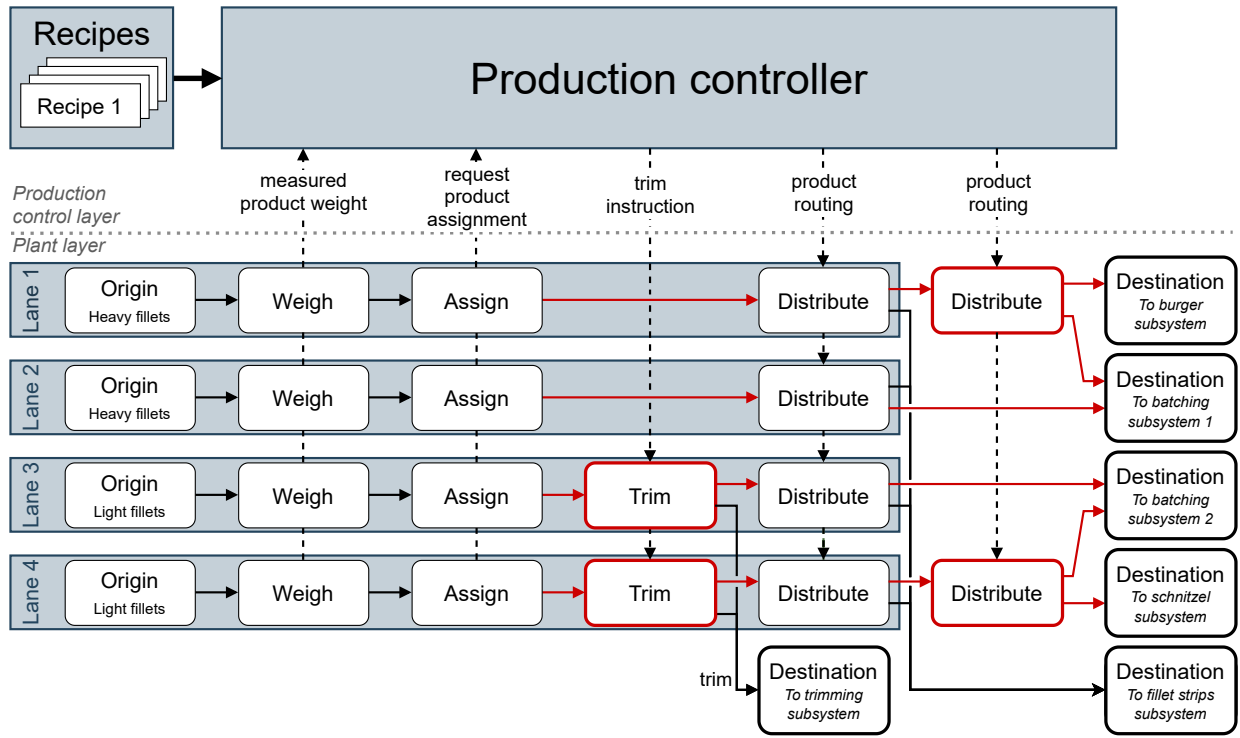

**Figure 1.** An example of the plant and production control layers of a fillet processing system.

### 2.1. Case study goal

In the case study a poultry fillet processing system is redesigned to minimize the mismatch between the inflow consisting of predominantly heavy fillets and production orders requiring mostly light fillets.

Each production order has its own specifications on what fillet weights are optimal for that order. For example, fillets between 250 ~ 350 grams are optimal for an order of chicken schnitzels. The performance of the fillet processing system is measured by how close the system is to reaching its throughput targets for the multiple production orders, over a variety of production scenarios. The goal of the case study is to optimize the functional connections in the facility layout of a poultry fillet processing system to optimize this performance.

### 2.2. System description

A fillet processing system can be decomposed into a (physical) plant layer and a (cyber) production control layer. In this subsection these two layers will be explained. An example of a fillet processing system is shown in Figure 1.

#### 2.2.1. Plant layer

The poultry fillet processing system is a subsystem in a poultry processing plant. In this subsystem the fillets are transported using conveyor belts, and operations on the fillets occur in-line while the belt keeps moving. The plant layer can be decomposed into modules which describe the production steps executed on the products flowing through the system. The origin and destination modules represent the inflow from and outflow to other subsystems. The plant layer contains the following modules:

**Origin**  This is where fillets flow into the subsystem. These fillets have a random variation in their weight. Generally, there are multiple incoming fillet flows, each with a different weight distribution. This is because broilers are divided upstream in the processing plant based on their weight to optimize production yield (e.g. they are often divided in separate categories ranging from light to heavy).

**Weighing**  The fillets are then weighed. The weight information is sent to the production controller, which uses this weight information to decide the production strategy (more on this later).

**Assignment**  When the fillets arrive at an assignment point, the production controller assigns a destination and a trim instruction to the fillet based on the production strategy (a trim instruction describes if and how much of the fillet should be trimmed).

**Trimming (optional)**  Some processing lanes have an inline trimming station. A trimming station can trim a small piece of a fillet, depending on the trim instruction that it was assigned. Only if a lane has a trimming station will the production controller assign a trim instruction. The produced trim is sent to a special trim destination. In the case study there are two lanes which have trimming. The placement of these two modules is a design parameter.

**Distribution** The next step in production is distribution. Distributors can be used to diverge product flow in one of two directions. Merging of product flows is only possible when fillets are used for fillet strips destinations (due to technological constraints). Each lane requires at least one distributor with a connection to the 'fillet strip subsystem', this is the default destination for fillets that cannot be used for other end products. In this case study there are two distributors to be placed freely besides the four required distributors. The placement of these two distributors is a design parameter. For both of these two distributors, one of the outputs is connected to either the burger or schnitzel destination, while the other is connected to either of the batcher destinations.

**Destination** Finally, the products arrive at their destination. In a fillet processing subsystem the destinations are the starting points of subsequent subsystems. There are multiple types of subsequent subsystems, each with their own requirements for optimal fillet weight (e.g. batching, burgers, schnitzels, etc.). As said before, the 'fillet strip subsystem' destination is a special destination which is is the default destination. The trim destination is another special destination for the produced trim.

For this case study, it is assumed that the order of processes is always the same: *Origin* → *Weighing* → *Assignment* [→ *Trimming*] → *Distribution* → *Destination*. In reality, some different designs could be feasible (e.g. distribution earlier in the fillet processing system), but this is not in the scope of this case study.

In the case study, the placement and functional connections of the two trimming modules and the two remaining distributors are optimized. These modules and their connections are highlighted in red in Figure 1.

### 2.2.2. *Production control layer*

The production controller is responsible for routing products through the system. Its goal is to fulfill the target throughput for all of the system's recipes. A *recipe* is an instruction of how products should be processed to meet production orders. A recipe describes:

- The product destination of products for this recipe.
- The priority of the recipe.
- The target throughput in fillets/minute.
- A lower and upper limit for fillet weight (post-trim).
- A weight limit for trimming (most customers want to limit how much of a fillet is trimmed).

The interactions between the production control layer and the plant layer of a fillet processing system are shown in Figure 1. The production controller functions as follows:

1. The production controller receives the fillet weight data

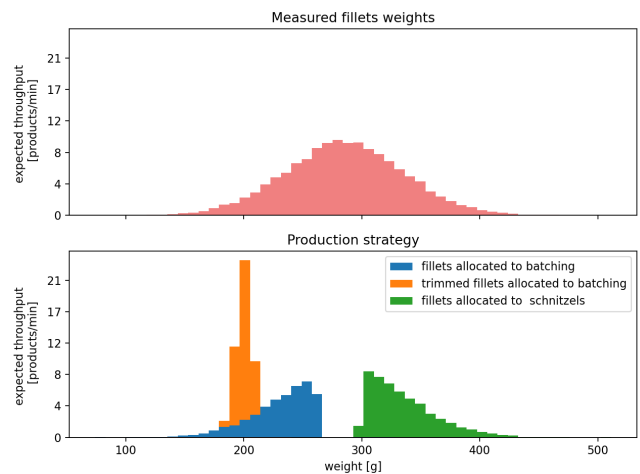

**Figure 2**. An example of the calculated production strategy for one of the lanes. It shows the measured throughput for different fillet weights, and the corresponding allocation to recipes. As can be seen in the bottom histogram, fillets around 280 gram are trimmed down to around 200 gram to better fit the production order.

from the weighing modules, and it collects the weight information for each lane.

2. The production controller uses a window of the last $N$ measured weights in each lane to build a histogram that predicts the expected throughput for different fillet weights (the bins of the histogram can be assigned to specific recipes).

3. The production controller calculates the production control strategies per lane every $t$ seconds using Algorithm 1. Figure 2 shows an example of such a strategy.

4. When a product arrives at an assignment module, the production controller assigns that product a routing to its destination and (if applicable) a trim instruction, based on the strategy for that lane and the measured weight of that product.

5. Whenever a product arrives at a distribution or trimming module, the destination or trim instruction is sent to that module.

## 3. Proposed method

In this section, the proposed method for design space exploration as shown in Figure 3 will be explained. The goal of the method is to automate the process of using discrete-event simulation for design space exploration of production systems. The output of the method is a set of recommended designs (there can be multiple recommendations). The four steps of the method, *explore design space*, *construct model*, *simulate*, and *evaluate* will be explained, along with the inputs and outputs for each step. Examples from the fillet processing case study described in Section 2 are used to illustrate these steps, however, the use of the method is not limited to fillet processing systems, but should extend to production systems in general.

---

**Algorithm 1:** Production control strategy calculation

```
 1  repeat
 2      Select (next) recipe with highest priority.
 3      Select lanes with a route to destination of recipe.
 4      Starting from lower weight limit of recipe:
 5      repeat
 6          Increase the weight range.
 7          Predict throughput for selected lanes.
 8      until (target throughput is reached) OR (recipe's upper
             weight limit is reached)
 9      if target throughput is not yet reached then
10          Identify selected lanes with trimming.
11          Starting from upper weight limit of recipe:
12          repeat
13              Increase the weight range for trimming
14          until (target throughput is reached) OR (recipe's
                 trimming weight limit is reached)
15      Assign (available) weight bins to selected lanes.
16      Make weight bins unavailable for next recipes.
17  until All recipes are processed
```

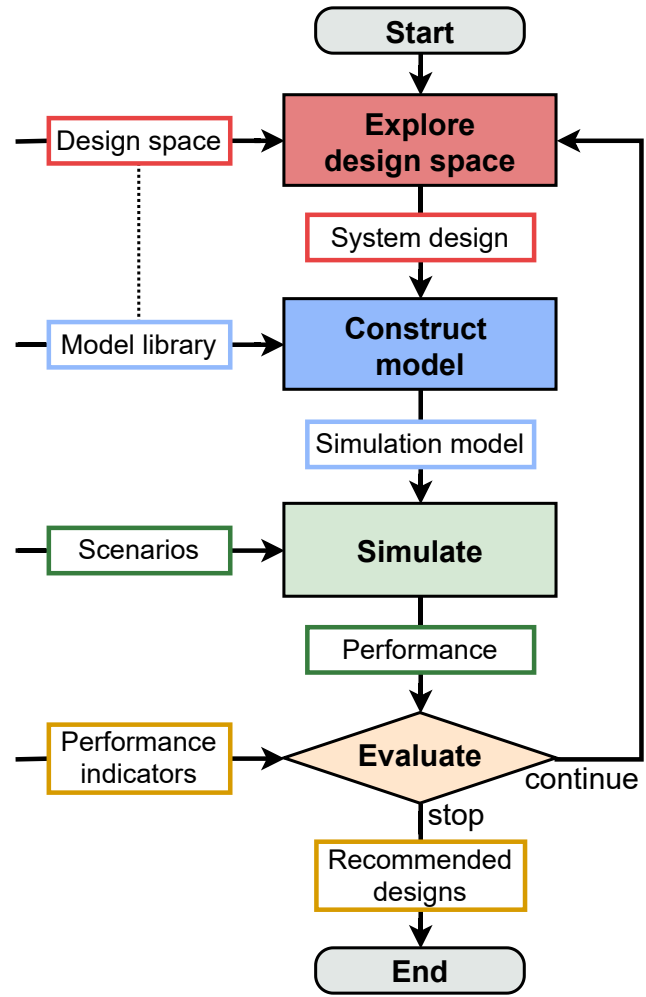

**Figure 3.** The method for design space exploration.

### 3.1. Design space exploration

To explore the design space of a production system, the design space must first be defined. In this method, the design space is defined through a Design Space Matrix, which describes all connections allowed in the design space. The Design Space Matrix represents all the different configurations of the system design. As shown in Figure 1, every module has a number of inflow and outflow ports. An element $(i, j)$ in the Design Space Matrix has a boolean value which denotes if module outflow port $i$ can have a connection with module inflow port $j$.

The design space exploration process starts by iterating through the possible configurations of the Design Space Matrix. Every iteration, a new design is created by, starting from the outputs of the origin modules, selecting one of the available connections in the Design Space Matrix. This step is then repeated with the outputs of those modules, until no more new connections can be made. It is possible that some modules are not connected in the generated configuration of system design. This makes it possible to evaluate the performance of a system with and without a specific module. However, either all or none of the ports of a module need to be connected. Configurations of the system design for which this is not the case are disregarded.

*Example*

A part of the design space for the case study is shown in Figure 4. In the figure, an example of a possible system design is highlighted. In the fillet processing system the parameters of the design space are: in which lanes the two trimming modules are located, and how the final two distributors are connected (the distributors which are not connected to the fillet strip destination). The connections of the system which cannot be altered only have one allowed connection in their respective rows of the Design Space Matrix. The case study design space is represented by a Design Space Matrix of $28 \times 28$, from which 1152 possible configurations of the system design can be derived (6 permutations for which lanes have trimmers, 6 permutations for which lanes have an extra distributor, and 32 permutations on how distributors can be connected to the destinations).

### 3.2. Model construction

The next step in this method is the 'Construct model' step, in which a simulation model of the system design is constructed using a model library in a simulation and modelling environment. This requires that all system modules have a corresponding model component (Figure 3 shows this relation between the design space and the model library). It also requires that these model components are modelled modularly, and that the model components can be connected dynamically. Finally, model construction requires that the production control layer of the system can

**Figure 4.** A part of the design space of the fillet processing system, the ticks show which connections are possible. An example of a possible system design is displayed by highlighting the selected connections (these connections are encircled)

be automatically adjusted to different system designs.

When constructing the model library, it is important to define if and what performance measures the model components should give as output after simulation. These performance measures are later used in the 'evaluate' step to score the system design on the performance indicators selected by the user of the design exploration method.

*Example*
For the case study, a model library was built in Anylogic. The model library is shown in Figure 5, it consists of all model blocks for the plant modules such as *weighing* and *distribution* described in Subsection 2.2.1. The model components for the destinations give as simulation output the percentages of target throughput reached for the recipes which are processed at these destinations. These model blocks are built using Anylogic's process-modelling library. To construct the model from these model components, Anylogic allows these model components to be connected dynamically. Figure 6 shows the part of the simulation model constructed based on the (partial) system design highlighted in Figure 4.

There is also one model component for the production controller of the fillet processing system. This component calculates the production strategy based on the measured product weights, and assigns a routing and (if necessary) trim instruction to products. To correctly set up the production control layer, the possible routings through the system have to be calculated for this specific system design. These routings can be deduced from the connections in the System Design Matrix, and from knowledge on how fillets flow through the modules. For example, for the model in Figure 6, it can be deduced that fillets can be routed to either the *schnitzel* or *filletStrips* destinations, and that these fillet can be trimmed if necessary.

### 3.3. Simulation

The third step of the method is *simulation*, in which the constructed model is simulated. Production systems often operate under a range of conditions. So, to accurately predict the performance of a production system requires that its behaviour is predicted for a selection of production

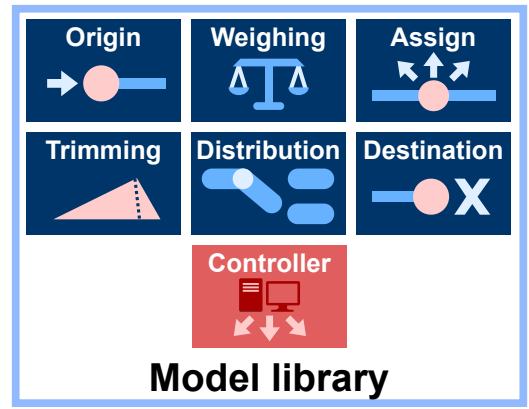

**Figure 5.** The model library used to construct the simulation models.

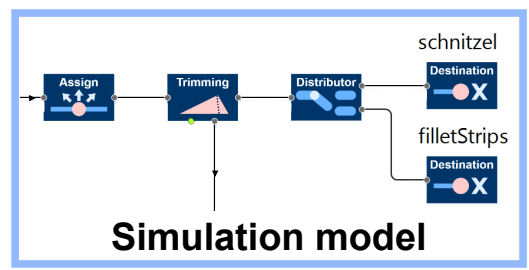

**Figure 6.** A part of the simulation model built in the construct model step, for which the system design highlighted in Figure 4 was the input. The assign module and trimming modules are connected to modules which are not depicted in this figure.

*scenarios*, similarly to Kikolski (2017). A *scenario* describes the conditions under which the production system operates.

What scenarios are to be simulated is up to the designer of the system. In very rigid production systems in which production is always the same, one scenario could be enough to predict the performance of the system. In a flexible manufacturing system many different production scenarios could be required to get a good picture of its performance. Alternatively, there could be scenarios for introduction of new products to the production line, or for the breakdown of machines.

The output of the simulation step is the performance of the system in the given scenarios. The relevant performance measures were chosen when constructing the model library.

*Example*
The poultry fillet processing system in this case study has two factors that influence its performance, both of which need to be taken into account to assess the performance of a system design. The first is that a poultry processing plant processes multiple different flocks of chicken during the day, and each flock has its own characteristic weight distribution. The weight distribution of the flock has a substantial influence on how well production targets can be met in the fillet processing system. The second is caused by the difference in production orders. A poultry processing plant generally produces multiple different production orders throughout the day, and production also changes throughout the year. For example, production is often focused on barbecue products during the summer. Depending on the production orders, the recipes of the fillet processing system are set. A good design for a fillet processing system requires that it performs well for many different recipes.

In the example case study, two different scenarios are evaluated. In the first production scenario, the focus is on sending fillets to the batching subsystems to produce batches of fillets. In the second production scenario, the focus is on sending fillets to the Burger and Schnitzel subsystems. The recipes for the two scenarios can be seen in Figure 7. In both instances, recipe 5 is used as the default recipe for fillets that do not fit any of the other recipes. For a case study on an actual system the number of scenarios needed to get a good impression of the systems performance would be higher.

For both scenarios, real-world weight data from the same flock was used. All simulation were done for 1 hour of production, resulting in a total of around 13000 fillets being processed. Simulation of all 1152 design configuration times 2 scenarios took 55 minutes (~500 million simulation steps), using a PC with an Intel(R) Core(TM) i5-8365U CPU with 16GB of RAM.

### Scenario 1

| Recipe | Destination | Priority | Target throughput [fillets / min] | Min. fillet weight [g] | Max. fillet weight [g] | Max. trim weight [g] |
|---|---|---|---|---|---|---|
| 1 | Batching 1 | 1 | 60 | 100 | 200 | 50 |
| 2 | Batching 2 | 2 | 60 | 150 | 200 | 100 |
| 3 | Burger | 3 | 30 | 200 | 300 | 100 |
| 4 | Schnitzel | 4 | 30 | 250 | 350 | 50 |
| 5 | Fillet strips | * | * | 0 | 1000 | 0 |

### Scenario 2

| Recipe | Destination | Priority | Target throughput [fillets / min] | Min. fillet weight [g] | Max. fillet weight [g] | Max. trim weight [g] |
|---|---|---|---|---|---|---|
| 1 | Batching 1 | 3 | 30 | 100 | 200 | 50 |
| 2 | Batching 2 | 4 | 30 | 150 | 200 | 100 |
| 3 | Burger | 1 | 60 | 200 | 300 | 100 |
| 4 | Schnitzel | 2 | 60 | 250 | 350 | 50 |
| 5 | Fillet strips | * | * | 0 | 1000 | 0 |

**Scenarios**

**Figure 7.** The two scenarios used in the case study. The differences between the two scenarios are highlighted in red.

### 3.4. Evaluation

The final step of the proposed design space exploration method is evaluation. During this step the performance of the system is evaluated. The performance of the system is evaluated on the basis of performance indicators which are chosen by the system designer. Which performance indicators are relevant depends for a big part on the chosen production scenarios (and different performance indicators might be required for different scenarios). Only the performance measures included when constructing the model library can be included in the calculation of the performance indicators.

After a design is evaluated, the result is stored and the next iteration of the method starts. This continues until either all designs are evaluated or until a stop condition is reached (for example, in case the designer is only interested in the first design that meets the specified requirements). Finally, the method yields one or more recommended designs, based on how they score on the chosen performance indicators.

*Example*
The fillet processing system has the same performance indicator for both of the two scenarios, which is the average percentage of the target throughput achieved for the four main recipes (not including the fillet strips recipe). If required, it would also be possible to choose separate performance indicators for different scenarios.

For this case study we are interested in all system designs that are Pareto optimal with respect to these two performance indicators. A design is Pareto optimal if no other design scores better in at least one performance indicator, without having to sacrifice in another performance indicator (Xu et al., 2015). Figure 8 shows a dot for the performance of every system design, with the performance in Scenario 1 and Scenario 2 on the horizontal and vertical axes, respectively. The same figure shows the 20 designs which are Pareto optimal, of which 5 designs are distinct (many designs are functionally equivalent. For example, the trimming modules can swap places without effecting the functionality and performance of the system). One of these Pareto optimal designs – highlighted in Figure 8 with a red star – is shown in Figure 9.

## 4. Discussion

In this section each of the four steps in the design space exploration method are discussed.

### 4.1. Discussion on design space exploration

The Design Space Matrix is not always detailed enough to describe the entire design space of a production system; many complex design specifications cannot be covered by it. An example of such a design specification could be: if A is connected to B, then C must be connected to D. Additionally, the design space can contain nonsensical or impermissible designs. For example, in the case study, if distributor 1 is connected to distributor 2, which in turn is connected to distributor 1 creating a loop. It is also possible that functionally equivalent designs are created. For the case study, trimming 1 and 2 are functionally equivalent, and so are distributors 1 and 2. As a result, only 1 in 4 designs is distinct. This could be solved by pruning the design space of functionally equivalent designs (Pimentel, 2017). Finally, the current implementation only allows to change connections, or to swap modules in or out. It does not allow for parameters of these modules (e.g. its processing time) to be changed. A solution to all of the listed problems could be to use a more descriptive method to specify the design space of the production system, for example; by using feature models such as in Vogel-Heuser et al. (2016).

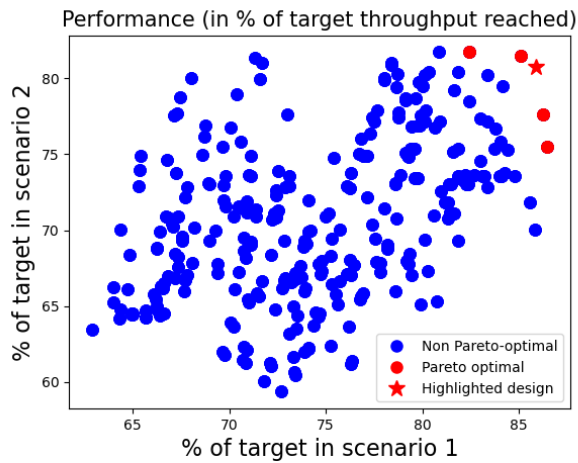

**Figure 8.** The performance in scenarios 1 and 2 of the 1152 different designs. Pareto optimal designs are shown in red. A model of the highlighted design is shown in Figure 9.

The current method for design space exploration – iterating through the different design configurations – is a rather naive approach. It would be interesting to investigate how the design space can be explored more intelligently. Pimentel (2017) lists some methods to explore the design space more intelligently such as simulated annealing and genetic algorithms.

### 4.2. Discussion on model construction

Building a model library requires a lot of expertise. However, the advantage of using a model library for model construction is its reusability. The model components can be reused for every different design, and even for future case studies in which similar performance measures are evaluated.

One of the challenges for model construction lies in constructing the model for the production control layer. For the fillet processing system, the production controller uses a generic design that can be used for any layout of the plant. This can be done because the production controller always interacts with the plant components in the same way. Only the production strategy needs to be adapted, which is automatically calculated based on the chosen recipes, and the possible product routings between assignment modules and destinations.

More research is required to asses how the proposed model construction method generalizes to other types of production systems in which the production control layer cannot be as easily adapted to different designs of the plant layer. For types of production systems in which the production control layer is always built up using a predefined rule set it might be possible to generate the model of the production control layer for every design of the plant layer. For types of production systems in which the production control layer is tailor-made to the design of the plant layer, the proposed model construction method might be entirely infeasible.

### 4.3. Discussion on simulation

The bottleneck of this method is in the simulation of all the different designs for all the various scenarios. The example case study required about an hour to simulate 2304 hours of production, spread over 1152 different designs, 2 scenarios each. However, the design space of the example case study is still fairly limited compared to a case study of an industrial size, especially when the goal is not to just optimize a fillet processing subsystem, but an entire poultry processing plant. When complexity increases, exploring the design space iteratively quickly becomes impossible. This is why it is so important to explore the design space more intelligently.

However, even when exploring the design space more intelligently, scalability can be an issue for much larger and more complex systems, as simulating even just a few design alternatives can be computationally expensive. Xu et al. (2015) notes that multi-fidelity simulation (simulating with models of different fidelity levels) could help bring the computational cost down by using lower fidelity models when there are too many possible design alternatives to simulate with high fidelity.

### 4.4. Discussion on evaluation

There are many different ways the evaluate step could be implemented. One method is simply to give all evaluated designs as output, along with the performances of the systems. This allows the designer to choose the preferred design out of the best available options. However, for more than three performance indicators, it becomes difficult to visualize the performance of the system, making it difficult for the designer to interpret the results and select the best option. Therefore, for more than three performance indicators, it would be recommended to use a different approach and let the method decide which designs are recommended.

However, setting up such a decision-process in itself is not necessarily straightforward, as it is difficult to com-pare the value of the different performance indicators. It becomes even more difficult when very different scenarios are evaluated. For example, there could be a scenario where the system operates as 'normal' and another scenario where one of the machines breaks down. One solution to this would be to set minimum requirements which a design must satisfy, e.g. a minimum of 80% of target throughput in normal conditions, and 60% in case of breakdown. Another solution could be to give weights to the different performance indicators, for example, a weight of 0.9 for performance under normal production and 0.1 for performance when there is a breakdown.

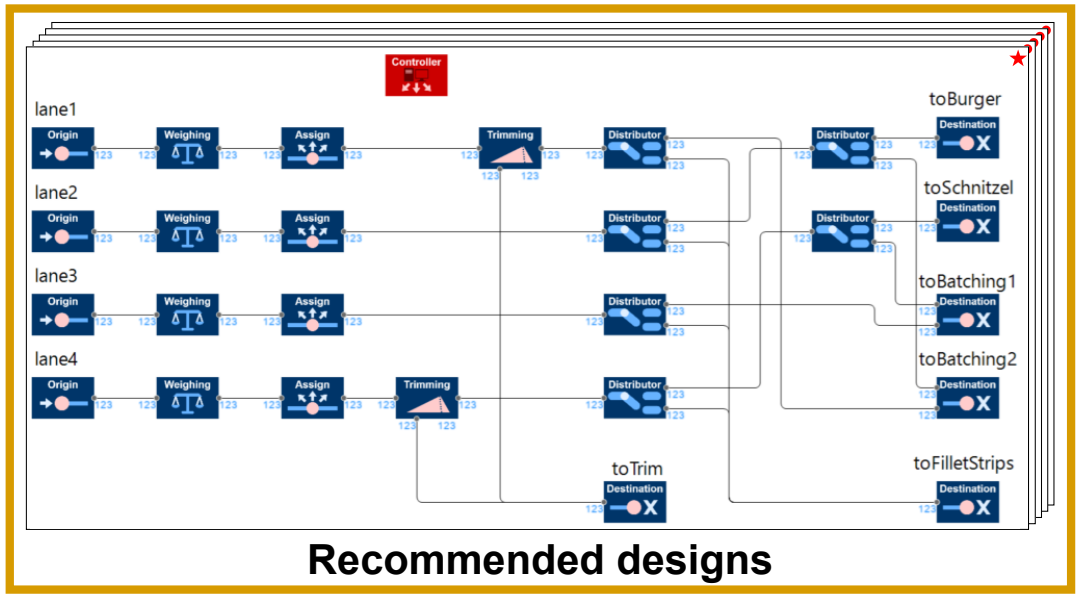

**Figure 9.** The design space exploration method outputs a set of recommended designs. The model of the design highlighted in Figure 8 is shown.

## 5. Concluding remarks

This paper presents a method for design space exploration using discrete-event simulation in which most of the steps are automated. This greatly reduces the time and effort required to iterate through different designs. Most of the effort when using this method is in constructing the model library. However, one of the biggest advantages is that this model library can be reused for future case studies.

The presented case study shows that this method can be effective for design space exploration of poultry processing systems. Our hypothesis is that these results extend to many other types of production systems.

One of the main challenges of this method is in dealing with case studies of increased complexity; the bottleneck of the method is in simulating the many different system designs for multiple scenarios. An improvement to the proposed design space exploration method would be to iterate through the designs more intelligently, which could be achieved by using optimization methods, and/or by using feedback from the 'evaluation' step to identify which direction design space exploration should continue.

If needed, the computational cost required for simulation could be reduced by using multi-fidelity simulation.